\documentstyle[12pt]{article}

\large
\newcommand{\be}{\begin{eqnarray}}
\newcommand{\ee}{\end{eqnarray}}

\begin{document}
\setlength{\baselineskip}{21pt}
\pagestyle{empty}
\vfill
\eject
\begin{flushright}
SUNY-NTG-92/45
\end{flushright}

\vskip 2.0cm
\centerline{\bf Random matrix theory and
spectral sum rules for the Dirac operator in QCD\footnote{Dedicated
to Hans Weidenm\"uller's 60th birthday.}}
\vskip 2.0 cm
\centerline{E.V. Shuryak and J.J.M.
Verbaarschot}
\vskip .2cm
\centerline{Department of Physics}
\centerline{SUNY, Stony Brook, New York 11794}
\vskip 2cm

\centerline{\bf Abstract}
We construct a random matrix model that, in the large $N$ limit,
reduces to the low energy limit of the QCD partition function
put forward by Leutwyler and Smilga. This equivalence holds for an
arbitrary number of flavors and any value of the QCD vacuum angle.
In this model, moments of the inverse squares
of the eigenvalues of the Dirac operator obey sum rules, which we
conjecture to be universal. In other words, the validity of the sum rules
depends only on the symmetries of the theory but not on its details.
To illustrate this point we show that the sum rules hold for
an interacting liquid of instantons.
The physical interpretation is that the way the thermodynamic limit
of the spectral density near zero is approached is universal. However,
its value, $i.e.$ the chiral condensate, is not.

\vfill
\noindent
\begin{flushleft}
SUNY-NTG-92/45\\
December 1992
\end{flushleft}
\eject
\pagestyle{plain}

\vskip 1.5cm
\renewcommand{\theequation}{1.\arabic{equation}}
\setcounter{equation}{0}
\noindent
\centerline{\bf 1. Introduction}

Since our main understanding of nonperturbative phenomena
in QCD comes from numerical simulations analytical results
in this direction are most welcome \cite{VANBAAL-1991,LEUTWYLER-SMILGA-1992A}.
In particular the issue of quark confinement has not been resolved.
What is better known \cite{THOOFT-1986}
is the mechanism of chiral symmetry breaking.
In the chiral limit the QCD partition function is invariant under
$SU(N_f)\times SU(N_f) \times U(1)$, whereas the axial $U(1)$ symmetry
is broken explicitly by quantum fluctuations. For more than one flavor
$(N_f > 1)$ the chiral symmetry is spontaneously broken down to
$SU(N_f) \times U(1)$ by the formation of quark condensates. According
to Goldstone's theorem this leads to $N_f^2 -1$ massless excitations. At low
energy, these are the relevant degrees of freedom of the QCD partition
function, and it is possible to write down an effective Lagrangian for
an arbitrary value of the vacuum angle $\theta$ in terms of
these degrees of freedom only
\cite{DIVECCHIA-VENEZIANO-1980,GASSER-LEUTWYLER-1987,LEUTWYLER-SMILGA-1992A}.
The partition function can be simplified further
by suppressing the space-time dependence
of the Goldstone modes altogether.

Recently, Leutwyler and Smilga \cite{LEUTWYLER-SMILGA-1992A} evaluated
the QCD partition function in this limit. This enabled them to write
down sum rules for the spectrum of
the Dirac operator. What enters in these sum rules is what we call
the microscopic spectral density as opposed to the continuum spectral
density that enters in the calculation of the condensate
\cite{BANKS-CASHER-1980}. The latter one is related to the thermodynamic
limit of the spectral density, whereas the former one provides information
on the way the thermodynamic limit is approached.
The question we would like to address is in how
far the sum rules are specific for the QCD partition function, and in
what sense they are universal.
Since the partition function of \cite{LEUTWYLER-SMILGA-1992A} only involves
constant fields one would suspect that the detailed structure of the
QCD vacuum is not important, and that the sum rules only depend
on the symmetries of the theory.

This adagio is taken to the extreme by constructing a model with
the correct chiral structure, but that apart from this does not contain
any other information. As is well known from random matrix theory,
the minimization of information leads to gaussian random matrix
ensembles \cite{BALIAN-1968}. The appropriate ensemble
will be constructed in section 2, and with the help of mathematical
techniques developed in the framework of Anderson localization
and compound nucleus scattering
\cite{EFETOV-1983,VERBAARSCHOT-WEIDENMUELLER-ZIRNBAUER-1985}
we are able to derive
the partition function that was obtained by Leutwyler and Smilga
(see section 3). Sum rules for one and two flavors are derived in
section 4. The structure of the random matrix model was inspired by
the semiclassical approximation to the QCD partition function where
instantons are the main degrees of freedom
\cite{DIAKONOV-PETROV-1986,SHURYAK-1988,NOWAK-VERBAARSCHOT-ZAHED-1989D}.
Therefore, it is natural to check the validity of these
sum rules for an instanton liquid model \cite{SHURYAK-VERBAARSCHOT-1993A}
of the QCD vacuum,
a task which is carried out
in section 5. Concluding remarks are made in section 6. Finally, in appendix
{\bf A} sum rules for a finite number of degrees of freedom are derived.

\vskip 1.5cm
\renewcommand{\theequation}{2.\arabic{equation}}
\setcounter{equation}{0}
\centerline{\bf 2. Random matrix model}
\vskip 0.5 cm
The Euclidean QCD partition function for $N_f$ flavors
and nonzero vacuum angle $\theta$ can be written as
\be
Z_{\rm QCD} = \sum_\nu e^{i\theta \nu}
< \prod^{N_f}_{f=1}\prod_n{'}(\lambda_n^2 + m_f^2) m_f^{\nu} >_{S_\nu(A)},
\ee
where the product is over the positive eigenvalues of the Dirac operator,
and the average $<\cdots>_{S_\nu(A)}$ is over gauge field configurations
with topological quantum number $\nu$ weighted by the gauge field action
$S_\nu(A)$. The topological part of the action, $\exp i\theta\nu$,
has been displayed explicitly.
Here and below, the mass matrix is taken to be diagonal.
According the the Atiyah-Singer theorem,
the Dirac equation in a gauge field with topological quantum number $\nu$
has exactly $\nu$ zero eigenvalues.
However, configurations with zero
total topological charge may actually be composed of spatially well-separated
components with opposite topological charge. Such configurations will give
rise to almost zero modes and thus will play an important role in the
chiral dynamics of the QCD partition function. Our starting point is
that the chiral properties of the QCD partition function are
determined by zero modes and almost zero modes only.
Although it is not necessary for the sake of the argument, it is instructive
to think of the field configurations as a superposition of $N_+$ instantons
and $N_-$ anti-instantons. Each isolated instanton or anti-instanton
has exactly one fermionic zero mode
with a definite chirality  \cite{THOOFT-1976A}.
In total we have $\nu = N_+ - N_-$ exact
zero modes. At finite separation of instantons and anti-instantons
the remaining modes are no longer exact zero modes of the
Dirac equation and give rise to nonzero overlap matrix elements $T$
of the Dirac operator \cite{DIAKONOV-PETROV-1985B}.

A model describing the zero mode part of the QCD partition function with
$N$ zero modes or almost zero modes in the Euclidean volume $V$ is defined by
\be
Z(\theta) = \sum_N \mu(N)\sum_{N_+}
\left( \begin{array}{c} N\\ N_+ \end{array} \right) e^{i\theta(N_+-N_-)}
\int {\cal D}T P(T)\prod_f^{N_f}\det \left (
\begin{array}{cc} m_f & iT\\
                 iT^\dagger & m_f
\end{array} \right ).
\ee
Zero modes of each chirality are treated as independently distributed identical
particles, hence the binomial factor. The distribution function
$\mu(N)$ of the total number of zero modes $N=N_+ + N_-$ is
peaked at some average value $\overline{N}$
with a width that is much smaller than
$\overline{N}$.
As long as a large number of different values of $N$ contribute
to the partition function with
roughly equal probability\footnote{If
$\mu(N)\sim\delta(N-\overline{N})$ the parity of $N_+ -N_-$ is the same as the
parity of $\overline{N} = N_+ + N_-$,
and the partition function has the additional symmetry
$Z(\theta = 0) = Z(\theta=\pi)$.}, our results do not
depend on the detailed shape of $\mu(N)$.
The average over all gauge field configurations in eq. (2.1)
is replaced by an average over gaussian distributed overlap matrix elements
with distribution function given by
\be
P(T) = \exp(-\frac N{2\lambda^2} T T^{\dagger}).
\ee
The integration measure ${\cal D}T$ is the Haar measure.
The structure of the overlap matrix, with off-diagonal blocks
$T$ and $T^\dagger$ and
diagonal blocks equal to the quark masses $m_f$ times the identity,
is dictated by
the chirality of the zero modes. The matrix $T$ is a $N_+\times N_-$
matrix, and for zero quark masses, the total overlap matrix has $|N_+ -N_-|$
exact zero eigenvalues.
The density $\overline{N}/V$ of the total number of zero modes
is kept fixed. As in the instanton liquid approximation
to the QCD partition function \cite{SHURYAK-1982}, it
is considered to be an external parameter. The thermodynamic limit can thus be
taken by letting $N\rightarrow\infty$, where here and below we will omit
the bar.
A similar model with $N_+ = N_-$ was considered
in \cite{NOWAK-VERBAARSCHOT-ZAHED-1989A,SIMINOV-1991}.

 The order parameter in the study of chiral symmetry breaking
is the quark condensate $<\bar q_f q_f>$ defined by
\be
<\bar q_f q_f>= \lim_{m_f\rightarrow 0}
\lim_{N\rightarrow \infty}-\frac {1}N \frac d{d m_f} \log Z(\theta),
\ee
where the order of the limits should be taken as indicated in the formula.
In general, $<\bar q_f q_f>$ depends on $\theta$. Its value at $\theta = 0$ in
the limit where for all flavors
$m_f \rightarrow 0$ is approached from above is denoted by
$-\Sigma$. By writing the determinant as the product
$\prod'(\lambda^2_n + m^2_f)$
one obtains the Banks-Casher formula \cite{BANKS-CASHER-1980} for $\Sigma$
\be
\Sigma = \pi<\rho_C(0)>_{Z(\theta = 0)},
\ee
where the average $<\cdots>_{Z(\theta)}$ is with respect to
the partition function
(2.2) (or (2.1) in the case of QCD).
The continuum spectral density is defined by
\be
\rho_C(\lambda) =\lim_{{\rm all} \,\,m_f\downarrow 0}
 \lim_{N \rightarrow \infty}\frac 1N \rho(\lambda),
\ee
and the spectral density $\rho(\lambda)$
is
\be
\rho(\lambda) = \sum \delta(\lambda -\lambda_n),
\ee
where the eigenvalues $\pm \lambda_n$ are the nonzero eigenvalues of the
overlap
matrix in the chiral limit.

We will also consider a different limit of
the derivative with respect of $m$ of the partition function, namely,
\be
\lim_{\begin{array}{c} ^{N\rightarrow \infty}\\ ^{Nm_f \,\,{\rm fixed}}
\end{array}} \frac 1{N^p} \frac {d^p}{dm^p_f}
\log Z(\theta).
\ee
Let us consider the case $p = 1$ in more detail.
Again writing the determinant as a product over eigenvalues, one
finds
\be
\lim_{\begin{array}{c} ^{ N\rightarrow \infty} \\
                        ^{Nm_f \,\,{\rm fixed}}
     \end{array} }
\left<\sum_{n>0} \frac {Nm_f}{\lambda^2_n N^2+ m^2_f N^2}\right>_{Z(\theta)} =
\lim_{\begin{array}{c} ^{N\rightarrow \infty} \\ ^{Nm_f \,\,{\rm fixed}}
\end{array}} \int_0^\infty
dx \left< \frac 1N \rho(\frac xN) \right>_{Z(\theta)}\frac {Nm_f}
{x^2 + m^2_f N^2}.\nonumber\\
\ee
What enters in this expression is what we call
the microscopic spectral density defined by
\be
\rho_S(x) = \lim_{\begin{array}{c}  ^{ N\rightarrow \infty} \\
{ ^{Nm_f} \,\,^{\rm fixed}}    \end{array} }
\frac 1N \rho(\frac xN),
\ee
as opposed to the continuum spectral density defined by in eq. (2.6). In this
limit, the spectral density function near zero is
enlarged proportional to the size (given by $N$) of the system.
Note that $\rho_S(x)$ depends on $Nm_f$.

\vskip 1.5cm
\renewcommand{\theequation}{3.\arabic{equation}}
\setcounter{equation}{0}
\centerline{\bf 3. Calculation of the partition function}
\vskip 0.5 cm
In order to evaluate the partition function (2.2)
the determinant is written as an integral over Grassmann
variables
\be
\prod_f\det \left ( \begin{array}{cc} m_f &i T
\\ iT^\dagger & m_f \end{array}\right)
= \int \prod_f {\cal D}\psi^f {\cal D}\phi^f \exp\sum_f
\left ( \begin{array}{c} \psi^{f*} \\ \phi^{f*} \end{array}\right)
\left ( \begin{array}{cc} m_f & iT \\ iT^\dagger & m_f \end{array}\right)
\left ( \begin{array}{c} \psi^f \\ \phi^f \end{array}\right),\nonumber\\
\ee
where the measure of the Grassmann integration is as usual
\be
{\cal D} \psi^f = \prod_i d\psi_i^f d\psi_i^{f*},
\ee
and the conjugation $^*$ is the  conjugation of the second kind
($i.e.$, $\psi^{**} = -\psi$) \cite{VERBAARSCHOT-WEIDENMUELLER-ZIRNBAUER-1985}.
The integral over $T$ is gaussian and can
be performed easily. In the partition function this results
in the factor
\be
\exp \frac{2\lambda^2}{N}
\psi^{f*}_i \psi_i^g \phi_j^{g*}\phi_j^f,
\ee
which represents a 4-fermion interaction.

The quartic term can be written as a sum of two squares
\be
\psi^{f*}_i \psi_i^g \phi_j^{g*}\phi_j^f =
\frac 14( \psi^{f*}_i \psi_i^g +  \phi_i^{f*}\phi_i^g)
(\psi^{g*}_j \psi_j^f + \phi_j^{g*}\phi_j^f )
-\frac 14( \psi^{f*}_i \psi_i^g -  \phi_i^{f*}\phi_i^g)
(\psi^{g*}_j \psi_j^f- \phi_j^{g*}\phi_j^f ).\nonumber\\
\ee
Each of the two squares can be linearized with the help of
a Hubbard-Stratonovich transformation
\cite{VERBAARSCHOT-WEIDENMUELLER-ZIRNBAUER-1985}. This allows us to perform
the Grassmann integrations at the expense of the introduction of the new
real valued integration variables $\sigma^{fg}$ and $\bar\sigma^{fg}$,
respectively. Apart from an irrelevant overall constant, the
partition function reduces to
\be
Z = \sum_N\mu(N)\sum_{N_+} \left( \begin{array}{c} N\\ N_+ \end{array} \right)
e^{i\theta(N_+-N_-)}
\int {\cal D}\sigma {\cal D}\bar\sigma &&{\det}^{N_+}(\sigma +i\bar\sigma
+m) {\det}^{N_-}(\sigma -i\bar\sigma+m)\nonumber\\ \times &&
\exp-\frac{N}{2\lambda^2}{\rm Tr} (\sigma+i\bar\sigma)(\sigma-i\bar\sigma).
\ee
As always, the measure of the integral over the
matrices $\sigma$ and $\bar \sigma$ is the
Haar measure. The diagonal mass matrix is denoted by $m$.

The complex matrix $\sigma + i\bar\sigma$ can be decomposed in 'polar
coordinates' as \cite{HUA-1963}
\be
\sigma + i\bar\sigma = U \Lambda V^{-1},
\ee
where $U$ and $V$ are unitary matrices and $\Lambda$ is a diagonal real
positive definite matrix. Since the r.h.s has $N_f$ more degrees of freedom
than the l.h.s., one has to impose constraints on the new integration
variables.
This can be achieved \cite{HUA-1963}
by restricting $U$ to the coset $U(N_f)/U(1)^{N_f}$, where
$U(1)^{N_f}$ is the diagonal subgroup of $U(N_f)$.
In terms of the new variables the partition function reads
\be
Z &=& \sum_N\mu(N)\sum_{N_+} \left( \begin{array}{c} N\\ N_+ \end{array}
\right)
\int J(\Lambda){\cal D}\Lambda {\cal D}U{\cal D} V \nonumber\\ &\times&
{\det}^{N_+}(U\Lambda V^{-1}+me^{i\theta/N_f})
{\det}^{N_-}(V\Lambda U^{-1}+me^{-i\theta/N_f})
\exp(-\frac{N}{2\lambda^2}{\rm Tr} \Lambda^2),
\ee
where the integral over $U$ is over $U(N_f)/U(1)^{N_f}$ and the integral
over $V$ is over $U(N_f)$.
A phase factor $\exp(-i\theta N_f)$ has been absorbed in the integration
over $V$.

For $N_f$ flavors we have $N_f$ condensates which break down the symmetry of
the action to $U(N_f)/U(1)^{N_f}$ leaving us with $N_f^2-1$ Goldstone
modes for $N_+ \ne N_-$. When the total topological charge is zero the phase
of the determinant also cancels which provides us with an additional zero mode.

The term proportional to $m$ plays the role of a small symmetry breaking
term. The integrals over the nonzero modes will
be performed by a saddle point integration at $m = 0$,
whereas the integrals over the
soft modes will be accounted for exactly at a fixed value of $mN$.
There are two types of nonzero modes, the phase $\exp i\alpha$
of the determinant $\det V U^{-1}$ for $N_+ \ne N_-$,
and the eigenvalues $\Lambda$. The
partition function  at $m = 0$ factorizes accordingly,
\be
Z(m=0) = \sum_N\mu(N)
\int d\alpha (\exp(i\alpha)+\exp(-i\alpha))^N\int J(\Lambda) d\Lambda
{\det}^N\Lambda \exp(-\frac{N}{2\lambda^2}{\rm Tr} \Lambda^2).\nonumber\\
\ee
The leading order contribution in $1/N$ of the integral over $\Lambda$
can be obtained by a saddle point
approximation. The saddle point equations for the $\Lambda$ integrals
read
\be
\Lambda_i = \pm \lambda.
\ee
The negative solution is not inside the integration manifold and can be
omitted.

The integral over $\alpha$, which ranges from $0$ until $2\pi$ ,
can be executed either before or after the summation over $N_+$. In the first
case we find zero for odd values of $N$ while for even values of $N$
only the term $N_+ = N_- = N/2$ contributes in which case the
$\alpha$ integral becomes soft. However,
in the second case the interference between contributions of all
different topologies results in the integrand $(2\cos\alpha)^N$ which
allows us to perform the integral by a saddle
point approximation in order to obtain the leading order contribution
in $1/N$. The saddle points are located at
$\alpha = 0$ and $\alpha = \pi$. It is at this point that
the distribution function $\mu(N)$ plays a role: the contribution of the
latter saddle point, $(-1)^N$, can be ignored after the summation over $N$,
whereas the contribution at $\alpha =0 $ yields a overall
factor $\mu(\overline{N})$ that does not contribute to the $m$ dependent
part of the partition function to be discussed below.

At the saddle points in $\Lambda$ and $\alpha$,
the $U-$dependence can be absorbed into $V$. The
the $U-$integration yields a finite irrelevant constant. Since the
phase $\exp i\alpha$ has already been extracted
the remaining integral over $V$ is over $SU(N_f)$ instead of $U(N_f)$.
We treat $m$ as a small parameter and expand the
determinants up to first order in $m$. At the saddle
point $\alpha = 0$ the result for $m$ dependent part of
the partition function is
\be
\frac{Z(m,\theta)}{Z(m=0,\theta)}= \int_{\det V = 1} {\cal D}V
\exp(\frac{N}{2\lambda}{\rm Tr}(m V^{-1} \exp(-i\theta/N_f)
+m V \exp(i\theta/N_f)),
\ee
which coincides with the result derived by Smilga and Leutwyler
\cite{LEUTWYLER-SMILGA-1992A} for the QCD partition function using chiral
perturbation theory.

The value of $\Sigma$ at $\theta = 0$ can be obtained from eq. (2.4)
\be
\Sigma(\theta =0) = \lim_{m\rightarrow 0}
\lim_{N\rightarrow \infty} \frac 1{2\lambda} \frac 1{N_f}
<{\rm Tr} (V+ V^{-1})>_{Z(\theta =0)}.
\ee
For $N_f = 1$ the integral over $V$ is absent, and
the sign of the quark condensate
is independent of the sign of $m$. For more than one flavor the order of the
limits allows us to perform
the $V$ integral by a saddle point approximation.
In the case of equal positive masses the saddle point for $\theta =0$
is at $V = {\bf 1}$ which allows us to identify the parameter $\lambda$
as
\be
\lambda = \frac{1}{\Sigma(\theta = 0)},
\ee
which completes the calculation of the partition function.

As observed in \cite{LEUTWYLER-SMILGA-1992A},
for more than one flavor the value of the condensate depends on the sign of
the quark mass. For example, in the case of two flavors with equal negative
masses and $\theta = 0$, the saddle point is at $V = {\bf -1}$.

It is also possible to introduce a complex mass in eq. (2.2) with
the mass in the lower block equal to the complex conjugate of the
mass in the upper block of the overlap matrix. In this case the final
result for the partition function depends only on the combination
$me^{i\theta}$ and its
complex conjugate which makes it possible \cite{LEUTWYLER-SMILGA-1992A}
to derive relations
between the $m$ and the $\theta$ derivatives of the partition function.
\vskip 1.5cm
\renewcommand{\theequation}{4.\arabic{equation}}
\setcounter{equation}{0}
\centerline{\bf 4. Sum rules for one and two flavors}
\vskip 0.5 cm
Sum rules for moments of the inverse squares of the eigenvalues of the Dirac
operator can be derived for an arbitrary number of flavors and for any
value of the total topological charge \cite{LEUTWYLER-SMILGA-1992A}.
In this section we only present a derivation for the simplest two cases of
one flavor and of two flavors with equal masses. Proofs of the
general results can be found in \cite{LEUTWYLER-SMILGA-1992A}.

For $N_f = 1$ there is no integration and the result for the ratio of
the massive and massless partition
function is particularly simple
\be
Z_{N_f=1}= \exp(Nm\Sigma\cos\theta),
\ee
which was first obtained in \cite{LEUTWYLER-SMILGA-1992A}.

In the case of two flavors the integral over $SU(2)$ can be performed
for an arbitrary mass matrix. Here, we restrict ourselves to the simpler case
of equal quark masses. In this case the ratio of the massive and the massless
partition function reduces to
\be
Z_{N_f = 2} = \int_0^{2\pi} \frac {d\phi}{\pi} \sin^2 \frac {\phi}{2}
\exp (2Nm\Sigma \cos\frac{\phi}2 \,\cos\frac{\theta}2),
\ee
where the factor $\sin^2 \frac {\phi}{2}$ results from the invariant
measure of $SU(2)$.
The integral is elementary and results in
\be
Z_{N_f = 2} = \frac{I_1(2 mN\Sigma \cos\frac{\theta}2)}
{mN\Sigma\cos\frac{\theta}2}.
\ee

The partition function can also be evaluated
by writing the fermion determinant
as a product over eigenvalues. For $N_f = 1$, we obtain
sum rule
\be
\exp(\Sigma mN \cos\theta) = <m^\nu
\prod_n{'}(1+\frac {m^2}{\lambda_n^2})>_{Z(m=0)},
\ee
where the average is with respect to the partition function $Z$ with
$m = 0$, which also includes the weight factor $\prod{'} \lambda_n^2$
involving the product of the nonzero eigenvalues of the Dirac operator.
The exclusion of zero eigenvalues in the product is denoted by a prime.
For two flavors a similar sum rule can be derived
\be
\prod_{x_{1n} \ge 0} (1+ \frac{4m^2N^2 \cos^2\frac{\theta}2}{x_{1n}^2 })
= <m^{2\nu} \prod_n{'}(1+\frac {m^2}{\lambda_n^2})^2>_{Z(m=0)}.
\ee
Here, the $x_{1n}$ is the $n$'th zero of the Bessel function $J_1$.
By expanding both the r.h.s. and the l.h.s. of eqs. (4.4) and (4.5)
in powers of $m$ we obtain sum rules for the inverse moments of the
eigenvalues $\lambda_n$. Additional sum rules for $N_f=2$ can be
derived from the expansion in powers of both $m_u$ and $m_d$
of the general expression for the partition function.

The sum rules (4.4) and (4.5) are for a fixed value of $\theta$. However,
we will investigate the validity of the Leutwyler-Smilga
sum rules in an interacting instanton vacuum where, for technical reasons,
the total topological is taken to be zero.
The corresponding partition function $Z_0$ is given by the Fourier transform of
$Z(\theta)$,
\be
Z_0 = \int_0^{2\pi} \frac{d\theta}{2\pi} Z(\theta).
\ee
which can be derived from the the decomposition
\be
Z = \sum_\nu e^{i\nu\theta} Z_\nu.
\ee
For one and two flavors, the integrals over $\theta$ are elementary. The
results
\be
Z_0 &=& I_0(Nm\Sigma) \quad {\rm for}\quad N_f = 1,\\
Z_0 &=& I_0^2(Nm\Sigma)-I_1^2(Nm\Sigma) \quad {\rm for}\quad N_f = 2,
\ee
were first derived by Leutwyler and Smilga \cite{LEUTWYLER-SMILGA-1992A}.
Sum rules are obtained
by equating (4.8) and (4.9) to the r.h.s. of (4.4) and (4.5)
restricted to zero topological charge, respectively.
For $N_f =1$ one finds
\be
I_0(Nm\Sigma) = <\prod_n(1+\frac {m^2}{\lambda_n^2})>_{\nu = 0},
\ee
and for $N_f =2$ we have
\be
I_0^2(Nm\Sigma)-I_1^2(Nm\Sigma) =
<\prod_n(1+\frac {m^2}{\lambda_n^2})^2>_{\nu = 0}.
\ee
The expansion of both sides of this equation in powers of $mN$ yields
sum rules for moments of the inverse eigenvalues of the Dirac operator.
The first two sum rules are
\be
<\sum_{n> 0} \frac 1{N^2\lambda^2_n}>_{\nu = 0} &=& \frac {\Sigma^2}{4N_f},\\
<(\sum_{n> 0} \frac 1{N^2\lambda^2_n})^2>_{\nu = 0} -
<\sum_{n> 0} \frac 1{N^2\lambda^2_n}>^2_{\nu = 0} &=& <\sum_{n> 0}
\frac 1{N^4\lambda^4_n}>_{\nu = 0}
-\frac {\Sigma^4}{16N_f^2(N_f+1)}.\nonumber\\
\ee
As was shown by Leutwyler and Smilga, both sum rules hold for arbitrary
$N_f$ and also be worked out for arbitrary topological charge. The
only modification is to replace $N_f$ by $N_f +|\nu|$.
In the case of two or more flavors additional sum rules can be derived by
varying the quarks masses independently \cite{LEUTWYLER-SMILGA-1992A}. We only
quote the result
\be
<\frac 1{N^4 \lambda_n^4}>_{\nu = 0} = \frac{\Sigma^4}{16N_f(N_f^2 -1)}.
\ee
Again, the result for arbitrary topological charge is obtained
by replacing  $N_f$ by $N_f +|\nu|$.
Formally, this sum rule diverges for $N_f=1$. That this is indeed the case
can be understood from the behavior of the spectral density at small
$\lambda$. For $N_+ = N_-$ we expect $\rho(\lambda) \sim
\lambda^{N_f+2}$, where
the factor $\lambda^{N_f}$ originates from the fermion
determinant and the  factor $\lambda^2$ from the Jacobian of the unitary
transformation that diagonalizes the overlap matrix.
The above sum rules diverge for $N_f = 0$. Consequently, we expect that in this
case, and therefore in the quenched approximation, that the thermodynamic
limit of the spectral density near zero is approached in a completely different
way.

The sum rules (4.12-14) can be expressed in the microscopic spectral density
by writing the sum over the eigenvalues as the integral
\be
\sum \frac 1{N^p\lambda_n^p} =  \int \frac{dx}{x^p}
{\rho_S(x)}_{|_{mN =0}}.
\ee
The second sum rule expresses an integral over the correlation function
$<\rho_S(x) \rho_S(x')>_{\nu =0}$ in terms of integrals over the average
microscopic spectral density $<\rho_S(x)>_{\nu=0}$.

Finally, let us write down a sum rule for nonzero values of $mN$. In the
case of one flavor one obtains from
the infinite product expansion for the Bessel function $I_0$,
\be
\sum_n \frac{mN\Sigma^2}{x_{0n}^2 + m^2N^2\Sigma^2} = \int_0^\infty dx
<{\rho_S(x)}_{|_{mN\,\, {\rm fixed}}}>_{\nu=0} \frac {mN}{x^2 + m^2N^2},
\ee
where $x_{0n}$ is the $n$'th zero of the Bessel function $J_0$.
One is tempted to invert this equality in order to obtain an analytical
answer for the microscopic spectral density. Since $\rho_S$ depends on
$mN$ there is no unique solution, but a trivial solution can be
written down readily,
\be
\rho_S(x) = \sum_n \delta(x-\frac{x_{0n}}{\Sigma}).
\ee
Because there is no reason to believe that the dispersion (4.13) is zero,
we do not expect that this is
the exact microscopic level density.

In appendix A it is shown that
for one flavor sum rules can also be derived at finite $N$.
These results provide an explicit proof that the thermodynamic limit
has been taken correctly.

\vskip 1.5cm
\renewcommand{\theequation}{5.\arabic{equation}}
\setcounter{equation}{0}
\centerline{\bf 5. Sum rules in the instanton liquid}
\vskip 0.5 cm
The partition function (2.1) can be approximated semiclassically by
a liquid of instantons. Instead of averaging over all gauge field
configurations, we average over the collective coordinates of the
instantons only, whereas 1-loop quantum fluctuations about the instantons
are included in the measure. The action in (2.1) is the instanton
action, which also includes the interaction between instantons.
We use the so called streamline \cite{YUNG-1988,VERBAARSCHOT-1991}
interaction supplemented by a core in order to stabilize the instanton
liquid. The fermion determinant is calculated in the space spanned by
the fermionic zero modes with overlap matrix elements that can be
derived from the streamline configuration
\cite{SHURYAK-VERBAARSCHOT-1992A}. More details on the above instanton
liquid model can be found in \cite{SHURYAK-VERBAARSCHOT-1993A}.

The numerical simulations were carried out for a liquid of 64 instantons
in a Euclidean space time volume of $(2.378)^3\times 4.756$ in units of
$\Lambda^{-4}$. Averages were obtained from 125 statistically
independent configurations.
Our main results are presented in Table 1. Calculations were done for one,
two and three massless flavors (see heading). The first row
shows the smallest eigenvalue and its average level motion (between brackets).
The condensate $\Sigma$ (row 2) is obtained from an extrapolation
of the spectral density to $\lambda = 0$. The results for the sum rule
(4.12) (row 3) are compared to the analytical result (row 4).
We find complete agreement inside the error bars.
The remaining rows involve numerical and theoretical values for
the quantities $S_1$ and $S_2$ defined by
\be
S_1 = \frac{<(\sum\frac 1{\lambda_n^2})^2>_{\nu = 0} -
            <\sum \frac 1{\lambda_n^4}>_{\nu=0}}
           {<\sum \frac 1{\lambda_n^2}>^2_{\nu=0}}, \qquad
S_2 = \frac{<(\sum\frac 1{\lambda_n^2})^2>_{\nu = 0}}
           {<\sum \frac 1{\lambda_n^2}>^2_{\nu=0}}.
\ee
Also in this case we find excellent agreement with the
theoretical values
\be
S_1^{\rm th} = \frac {N_f}{N_f+1}, \qquad
S_2^{\rm th} = \frac {N_f^2}{N_f^2-1},
\ee
where the derivation for $S_2$ does not hold for $N_f = 1$.
We consider this a strong argument in favor of the universality of the
sum rules.

\vskip 1.5cm
\renewcommand{\theequation}{6.\arabic{equation}}
\setcounter{equation}{0}
\centerline{\bf 6. Discussion and conclusions}
\vskip 0.5 cm

The main conclusion of this work is that the low energy chiral limit of
the QCD partition function as derived by Leutwyler and Smilga
coincides with a large $N$ limit
of a random matrix theory.
Therefore all conclusions of \cite{LEUTWYLER-SMILGA-1992A} also
pertain to the chiral random matrix theory proposed in section 2.
The random matrix theory resembles
the partition function of a gas of instantons. The main difference is
that instead of inheriting its disorder from the distribution of the
collective coordinates, the overlap matrix elements are taken to be
independently distributed according to a gaussian random matrix ensemble.

{}From random matrix theory it is know that \cite{BRODY-ETAL-1981}
microscopic level correlations
such as for example the nearest neighbor level spacing distribution
or the variance of the number of levels as a function of the average
number of levels in a given interval are universal. They only depend
on the symmetries of the Hamiltonian but not on the details
of the matrix elements.
The spectral sum rules put forward by Smilga and Leutwyler
involve the microscopic level density. Since they
can be obtained from a chiral random matrix theory
we expect that the sum rules are also universal,
and that they do not depend on the details of the low-energy structure
of the theory but only on its symmetries. In particular, they depend on
the number of flavors and, for example, diverge in the quenched
approximation. This conclusion
has been confirmed by simulations of an instanton liquid  with
overlap matrix elements that differ strongly from those
of a gaussian random matrix ensemble.
Inside the statistical uncertainty of the calculations
all sum rules were reproduced.
The physical interpretation is that the way the thermodynamic limit
of the spectral density near zero is approached is universal. However,
its value, $i.e.$ the chiral condensate, depends on the details of the
theory.

The advantage of a random matrix model is that it is amenable to powerful
mathematical techniques
\cite{MEHTA-1991,VERBAARSCHOT-WEIDENMUELLER-ZIRNBAUER-1985}
that makes it possible to obtain explicit analytical results for {\it e.g.}
the microscopic
spectral density, which, by virtue of the universality arguments discussed
above, also hold for the QCD partition function. Work in this direction is
under way.

\vglue 0.6cm
{\bf \noindent  Acknowledgements \hfil}
\vglue 0.4cm
 The reported work was partially supported by the US DOE grant
DE-FG-88ER40388. We acknowledge the NERSC at Lawrence Livermore where
most of the computations presented in this paper were performed.
We would like to thank A. Smilga, M.A. Nowak and I. Zahed for
useful discussions.

\vskip 1.5cm
\renewcommand{\theequation}{A.\arabic{equation}}
\setcounter{equation}{0}
\noindent
\centerline{\bf Appendix A}
\vskip 0.5 cm
In this appendix we derive the sum rules (4.12) and (4.13)
for one flavor without relying on
a saddle point approximation. For $N_+ = N_-= N/2$ and $N_f = 1$ the partition
function simplifies to
\be
Z = \int d\sigma d\bar\sigma (\sigma + i\bar\sigma -m)^{\frac N2}
(\sigma - i\bar\sigma -m)^{\frac N2}
\exp(-\frac{N}{2\lambda^2}(\sigma+ i\bar\sigma)(\sigma - i\bar\sigma)),
\ee
where an irrelevant overall constant has been suppressed. At finite $N$
the pre-exponential factors can be expanded as a binomial series which provides
us with an expansion in powers of $m$. The coefficients are elementary
integrals, and for the $m-$dependent part of the partition function we find
\be
\frac {Z(m)}{Z(0)} = 1 + \frac {m^2N^2 \Sigma^2}{4}
+ \frac {m^4 N^4\Sigma^4}{64}(1-\frac 2N) + \cdots.
\ee
It should be noted that no approximations have been made. On the other hand,
the fermion determinant can be written as a product over the eigenvalues
which leads to the expansion
\be
\frac {Z(m)}{Z(0)} = 1 + m^2 \left<\sum_n\frac 1{\lambda^2_n}\right >_{\nu=0}
 + m^4
\frac 12 \left <\sum_{n\ne n'} \frac 1{\lambda_n^2\lambda_{n'}^2}\right >_{\nu
= 0} + \cdots,
\ee
where the average is with respect to the massless partition function.
By equating the coefficients of the powers of $m^2$ we obtain sum rules
for the inverse powers of the eigenvalues that are valid for
any value of $N$. The sum rules (4.12) and (4.13) for $N_f =1$ are reproduced
by keeping only the leading order terms in $1/N$. We observe that the
first sum rule is valid for any value of $N$. The second sum rule
is modified by the factor $(1-2/N)$. For $N=2$ we find zero which is
correct because in this case there are no terms that contribute to
the sum $n'\ne n$ in (A.3).
\vfill
\eject

\newpage
\vskip 1cm
\begin{tabular}{||l|l|l|l||} \hline
    & $N_f = 1$ & $N_f = 2$  &  $N_f = 3$ \\
\hline
 $\lambda_1$          & 0.0255(98) & 0.0531(164) & 0.1078(335)  \\
 $\Sigma$             & 1.90(5)    & 1.33(5)     & 0.90(10)     \\
 $\sum 1/\lambda_n^2$ &  0.89(3)   &  0.21(1)    & 0.066(2)     \\
 $ \Sigma^2/4N_f$     &  0.90(3)   &  0.22(1)    & 0.068(7)     \\
 $ S_1$               &  0.53(3)   &  0.70(5)    & 0.83(7)      \\
 $ S_1^{\rm th}$      &  1/2       &  2/3        & 3/4          \\
 $ S_2$               &  2.0(4)    & 1.20(12)     & 1.24(12)     \\
 $ S_2^{\rm th}$      & $\infty$   &  4/3        & 9/8          \\
 \hline
\end{tabular}
\vskip 0.5 cm
\noindent
{\bf Table 1.} Numerical results for the spectral properties
of the Dirac operator in the gauge field of a liquid of instantons.
{}From the comparison of the 3rd and 4th, the 5th and 6th and the
7th and 8th rows we conclude that the Leutwyler-Smilga sum rules
are observed by this model. For the definition and further discussion
of the observable, we refer to section 5.

\end{document}